\documentclass[aps,pra,twocolumn,groupedaddress,showpacs]{revtex4}

\usepackage{graphicx}
\usepackage{dcolumn}
\usepackage{bm}

\begin{document}


\title{Rb$^{\ast}$He$_{n}$ exciplexes in solid $^{4}$He}



\author{A. Hofer}
\email[]{adrian.hofer@unifr.ch}
\author{P. Moroshkin}
\author{D. Nettels}
\author{S. Ulzega}
\author{A. Weis}

\affiliation{D\'{e}partement de Physique, Universit\'{e} de
Fribourg, Chemin du Mus\'{e}e 3, 1700 Fribourg, Switzerland}
\homepage[]{www.unifr.ch/physics/frap/}

\date{\today}

\begin{abstract}
We report the observation of emission spectra from
Rb$^{\ast}$He$_{n}$ exciplexes in solid $^{4}$He. Two different
excitation channels were experimentally identified, viz., exciplex
formation via laser excitation to the atomic 5$P_{3/2}$ and to the
5$P_{1/2}$ levels. While the former channel was observed before in
liquid helium, on helium nanodroplets and in helium gas by
different groups, the latter creation mechanism occurs only in
solid helium or in gaseous helium above 10 Kelvin. The
experimental results are compared to theoretical predictions based
on the extension of a model, used earlier by us for the
description of Cs$^{\ast}$He$_{n}$ exciplexes. We also report the
first observation of fluorescence from atomic rubidium in solid
helium, and discuss striking differences between the spectroscopic
feature of Rb-He and Cs-He systems.
\end{abstract}

\pacs{32.30.-r, 33.20.-t, 33.20.Ea, 67.40.Yv, 67.80.-s}

\maketitle


\section{Introduction}\label{sect.Introduction}
The formation process of alkali-He$_{n}$ exciplexes, i.e., of
bound states of an excited alkali atom with one or more ground
state helium atoms, was studied in recent years in superfluid
\cite{EmissionSpectraCsHeExcimersColdHeGas;Yabuzaki,{EmissionSpectraRbHeExciplexesColdHeGas;Yabuzaki}}
and in solid
\cite{DiscoveryOfDumbbellShapedCsHenExciplexes;Nettels} helium.
These studies have given support to earlier proposals
\cite{ExcitedPStatesOfAlkalisInLiquidHe;DupontRoc,PressureShiftOfAtomicResonanceLinesInLiquidSolidHe;Kanorsky},
which tentatively explained the quenching of atomic fluorescence
from light alkali atoms (Li, Na, K) in condensed helium by the
formation of alkali-helium exciplexes, whose emission spectra are
strongly red-shifted with respect to the atomic resonance lines.
Exciplex formation was also studied on the surface of helium
nanodroplets
\cite{SpectroscopyOfCsAttachedToHeliumNanodroplets;Stienkemeier,
FormationOfKHeExciplexesOnTheSurfaceOfHeliumNanodropletsStudiedInRealTime;Stienkemeier,
RbHeExciplexFormationOnHeliumNanodroplets;Ernst,
AlkaliHeliumExciplexFormationOnTheSurfaceOfHeliumNanodroplets1;Scoles,
AlkaliHeliumExciplexFormationOnTheSurfaceOfHeliumNanodroplets2;Scoles}
and in cold helium gas
\cite{EmissionSpectraOfAlkaliMetalHeExciplexes;Yabuzaki,
EmissionSpectraCsHeExcimersColdHeGas;Yabuzaki,
EmissionSpectraRbHeExciplexesColdHeGas;Yabuzaki}. Recently we have
performed an experimental and theoretical study of the
Cs$^{\ast}$He$_{n}$ exciplex formation process in the hcp and bcc
phases of solid $^{4}$He
\cite{CsHenExciplexesInSolidHe;Moroshkin}. A comparison with the
results of
\cite{{EmissionSpectraCsHeExcimersColdHeGas;Yabuzaki},{EmissionSpectraRbHeExciplexesColdHeGas;Yabuzaki}}
has revealed that the exciplex formation mechanism in solid helium
differs from the one in superfluid helium and in cold helium gas.
We concluded that exciplexes in solid helium result from the
collective motion of several nearby helium atoms which approach
the Rb atom simultaneously, while in liquid and gaseous helium the
binding of the helium atoms proceeds in a time sequential way.

The motivation for the present study of the Rb-He system arose
from the question whether the collective mechanism is specific for
Cs in solid helium, or whether it also holds for other alkali
atoms. While the light alkali atoms (Li, Na, K) do not emit
resonance fluorescence when excited in condensed helium, atomic
cesium fluoresces both in superfluid and in solid helium, when
excited on the D$_1$ transition. Rubidium represents an
intermediate case, as it was reported
\cite{PressureDependentQuenchingRb5PStatesLiquidHe;Yabuzaki} to
fluoresce in liquid helium when excited on the D$_1$ transition
with a yield which is strongly quenched with increasing He
pressure. No fluorescence from Rb in solid helium was observed in
the past, although it was shown that optically detected magnetic
resonance can be used to detect light absorption on its D$_1$
transition \cite{OptDetecNonradAlkaliAtoSolidHe;Eichler}.
\begin{figure}[h]
\includegraphics[width=8.5cm]{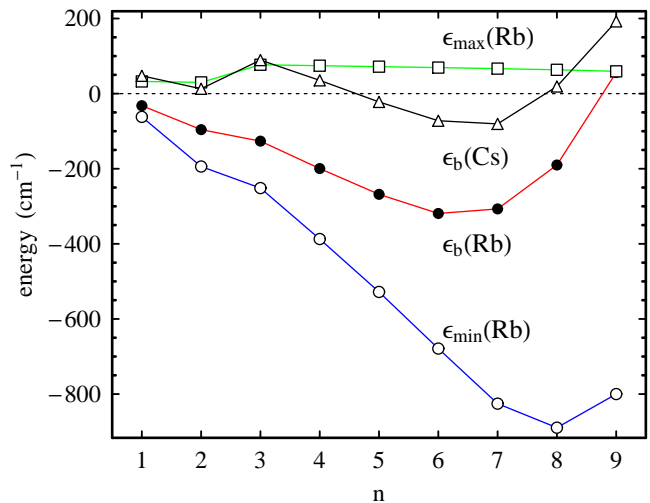}
\caption{Calculated energies of Rb(A$^{2}\Pi_{1/2}$)He$_{n}$
exciplexes as a function of the number $n$ of attached helium
atoms. All energies (defined in
Fig.\,\ref{fig:PotentialPlotRbHe2RbHe6}(b)) are given with respect
to the dissociation limit, i.e., the energy of the $5P_{1/2}$
state of free Rb . Shown here are the depths of the potential
wells $\epsilon_{min}$(Rb) (open circles), the barrier heights
$\epsilon_{max}$(Rb) (open squares) and the binding energies
$\epsilon_{b}$(Rb) (solid dots). The binding energies
$\epsilon_{b}$(Cs) (open
triangles) of Cs exciplexes from \cite{CsHenExciplexesInSolidHe;Moroshkin} are shown for comparison. \\
}\label{fig:EnergyLevel}
\end{figure}

A major difference between cesium and rubidium exciplexes
Rb/Cs(A$^{2}\Pi_{1/2}$)He$_{n}$ becomes apparent from
Fig.\,\ref{fig:EnergyLevel} which shows the calculated binding
energies $\epsilon_b$(Rb) ($\epsilon_b$(Cs)) of the exciplexes as
a function of the number $n$ of bound helium atoms for Rb (Cs).
For Cs only exciplexes with 5, 6 and 7 helium atoms have their
energy below the dissociation limit and are therefore stable,
while for Rb all exciplexes with $n=1\ldots8$ are stable.

For cesium the binding energy has a local minimum for $n=2$
(quasi-bound complex) and there is a potential barrier that
hinders the formation of exciplexes with more than 2 helium atoms
in  a sequential manner. As evidenced by the measurements of
\cite{EmissionSpectraCsHeExcimersColdHeGas;Yabuzaki} the
Cs$^{\ast}$He$_{n=2}$ exciplex is therefore the largest complex
that can be formed by a sequential attachment of He atoms. Larger
complexes can only be formed in a collective way, which becomes
possible in pressurized solid helium
 \cite{DiscoveryOfDumbbellShapedCsHenExciplexes;Nettels}. The
largest stable complex will be the one with the lowest binding
energy. For Rb all the exciplexes with $n=1\dots8$ are stable, so
once the Rb$^{\ast}$He$_{n=1}$ exciplex is created all larger
complexes can be formed with high probability by the sequential
filling of the helium ring until the state with the lowest binding
energy is reached. In helium environments with lower densities
than pressurized solid helium the time intervals between
successive attachments is long enough to permit the exciplex to
fluoresce, so that fluorescence from all intermediate exciplexes
Rb$^{\ast}$He$_{n=1\dots6}$ can be observed in gaseous helium
\cite{EmissionSpectraRbHeExciplexesColdHeGas;Yabuzaki}. The
results presented below indicate that in solid He the
Rb(A$^{2}\Pi_{1/2}$)He$_{n}$ formation process is so rapid that
any intermediate configurations have no time to emit fluorescence.
For Rb in solid helium one therefore expects that only the most
strongly bound Rb$^{\ast}$He$_{6}$ exciplex is formed.


In Sec.\,\ref{sect.Theory} we review the theoretical model for the
description of exciplex spectra developed in
\cite{CsHenExciplexesInSolidHe;Moroshkin} and extend it to the
Rb-He system. In Sec.\,\ref{sect.ExperimentalResults} we introduce
the experimental setup and present experimental emission and
excitation spectra of rubidium-helium exciplexes. In
Sect.\,\ref{sect.Discussion} we compare the experimental results
with the theoretical model calculations as well as other
experiments and discuss the different decay channels of excited Rb
in solid helium.

\section{Theory}\label{sect.Theory}
\begin{figure}[h]
\includegraphics[width=5.5cm]{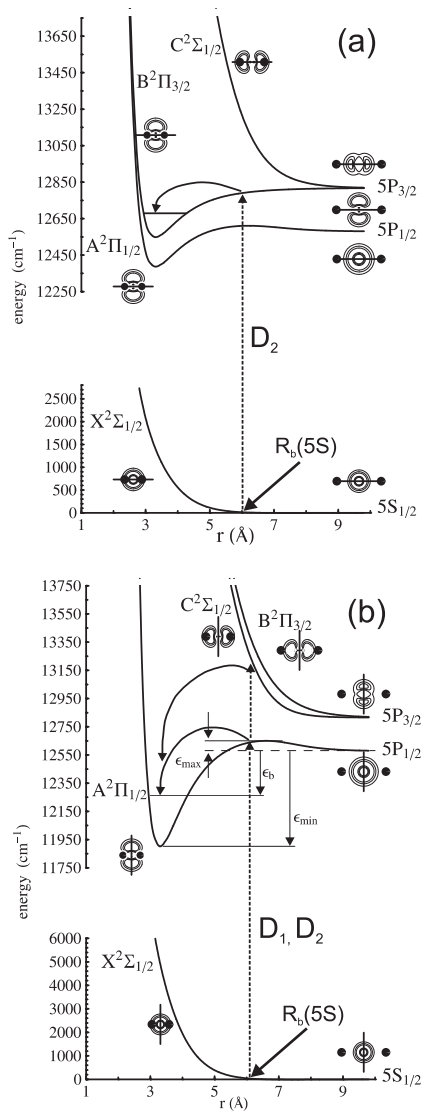}
\caption{Adiabatic potentials of the Rb$^{\ast}$He$_{n}$ system:
(a) Rb$^{\ast}$He$_{2}$ , (b) Rb$^{\ast}$He$_{6}$. The equilibrium
bubble radius of the ground state Rb atom is indicated with
R$_{b}$(5S). The energies shown in Fig.\,\ref{fig:EnergyLevel} as
a function of the number of bound helium atoms are visualized in
(b).} \label{fig:PotentialPlotRbHe2RbHe6}
\end{figure}

We briefly describe the theoretical approach of our calculation of
the Rb$^{\ast}$He$_{n}$ exciplex emission spectra for $n=1-9$. The
model used is an extension of the calculations performed earlier
for cesium-helium exciplexes
\cite{{CsHenExciplexesInSolidHe;Moroshkin},{DiscoveryOfDumbbellShapedCsHenExciplexes;Nettels}}
and we shall review only the basic principles and assumptions. We
consider only the interaction of the excited Rb atom with the $n$
helium atoms that form the exciplex and neglect the influence of
the helium bulk. The largest perturbation comes from the close
helium atoms that form the exciplex and it is therefore a good
approximation to neglect the helium bulk. The interaction between
the Rb atom and one ground state helium atom is described as a sum
over semi-empirical pair potentials
\cite{LDependentPseudoPotentialAlkaliHe;Pascale}
\begin{equation}
V^{\text{Rb-He}}_{n}(r)=\sum_{i=1}^{n}V^{5P}(\mathbf{r_{i}})\,,
\end{equation}
where $\mathbf{r_{i}}$ is the position of the i-the helium atom
with respect to the position of the Rb atom. After including the
spin-orbit interaction of the Rb valence electron and the
helium-helium interaction, $V_{n}^{\text{He-He}}(r)$, modeled as
the sum over interaction potentials
 \cite{AbInitioCalculationsForHelium;Aziz} between neighboring
helium atoms the total interaction Hamiltonian is given by
\begin{equation}
V_{\text{Rb}^{\ast}\text{He}_{n}}(r)=V^{\text{Rb-He}}_{n}(r)+V_{n}^{\text{He-He}}(r)+(2/3)\Delta\mathbf{L}\cdot\mathbf{S}, \label{eq:VRb-He}
\end{equation}
where $\Delta=237.6$\,cm$^{-1}$  is the fine structure splitting
of the rubidium $5P$ state in the free atom. $\mathbf{L}$ is the
orbital angular momentum operator and $\mathbf{S}$ the electronic
spin operator. Next, the total interaction operator
$V_{\text{Rb}^{\ast}\text{He}_{n}}(r)$ is represented in the basis
$|n,L,S\rangle$ and diagonalized algebraically. Exciplexes of two
different structures are formed as in the case of cesium-helium
exciplexes. When one or two helium atoms are bound the electronic
wavefunction has an apple shape with the helium atoms attached in
its dimples, whereas for $n>2$ the electronic wavefunction has a
dumbbell shape, with the bound helium atoms distributed along a
ring around the dumbbell's waist. The potential curves leading to
the formation of these two classes of structures are represented
in Fig.\,\ref{fig:PotentialPlotRbHe2RbHe6} using the examples of
Rb$^{\ast}$He$_{2}$ and Rb$^{\ast}$He$_{6}$. The potential curves
shown represent the $r$-dependent eigenvalues of the operator
$V_{\text{Rb}^{\ast}\text{He}_{n}}(r)$ of Eq.\,(\ref{eq:VRb-He}).
In
the same figures we also show the ground state potentials $nV_{\sigma}^{5S}(r)+V_{n}^{\text{He-He}}(r)$.\\

As can be seen from Fig.\,\ref{fig:PotentialPlotRbHe2RbHe6}(a) the
energetically most favorable formation channel for
Rb$^{\ast}$He$_{2}$ proceeds via D$_2$ excitation; when two helium
atoms approach along the nodal line of the apple-shaped electron
distribution of the B$^{2}\Pi_{3/2}$ state, they are attracted
into the potential minimum. When the system is excited on the
D$_{1}$ transition the approaching helium atoms see a repulsive
spherical electronic distribution of the Rb atom at large
distances with a potential barrier of 29$\,\text{cm}^{-1}$. We
recall that the corresponding barrier height in cesium is
79$\,\text{cm}^{-1}$ \cite{CsHenExciplexesInSolidHe;Moroshkin} due
to the larger spin-orbit interaction energy in that atom
\cite{ExcitedPStatesOfAlkalisInLiquidHe;DupontRoc}. The
approaching helium atoms deform the electronic configuration of
the $5P$ state from  spherical to apple shaped.

The exciplexes with $n>2$
[Fig.\,\ref{fig:PotentialPlotRbHe2RbHe6}(b)] have no potential
well in the B$^{2}\Pi_{3/2}$ state, which is purely repulsive and
which correlates to the $5P_{3/2}$ atomic state. However, the
A$^{2}\Pi_{1/2}$ state possesses a potential well and a potential
barrier. The barrier is associated with the transformation of the
electronic wavefunction from spherical to dumbbell-shaped when
several helium atoms approach the Rb atom. Exciplexes with $n>2$
can only be formed in the A$^{2}\Pi_{3/2}$ state.

The electronic distributions of the rubidium-helium system for the
different states at various interatomic separations are
illustrated by pictographs in
Fig.\,\ref{fig:PotentialPlotRbHe2RbHe6}. The solid lines represent
the quantization axis, which is the internuclear axis for
Rb$^{\ast}$He$_{n\leq 2}$ and the symmetry axis of the helium ring
for the Rb$^{\ast}$He$_{n>2}$ complexes, while helium atoms are
drawn as filled disks with a radius of 3.5 \AA.

In a next step we have calculated the vibrational zero-point
energies for all Rb$^{\ast}$He$_{n}$ for $n=1\ldots9$. Details of
this calculation were discussed in
\cite{CsHenExciplexesInSolidHe;Moroshkin} for the case of cesium.
Only the lowest vibrational state is considered as higher
vibrational states are not populated at the temperature (T=1.5 K)
of the experiment. The binding energies $\epsilon_{b}$(Rb),
$\epsilon_{b}$(Cs), the well depths $\epsilon_{min}$(Rb) and the
barrier heights $\epsilon_{max}$(Rb) are shown in
Fig.\,\ref{fig:EnergyLevel} for
Rb(A$^{2}\Pi_{1/2}$)He$_{n=1\ldots9}$.

\begin{figure}[h]
\includegraphics[width=6.5cm]{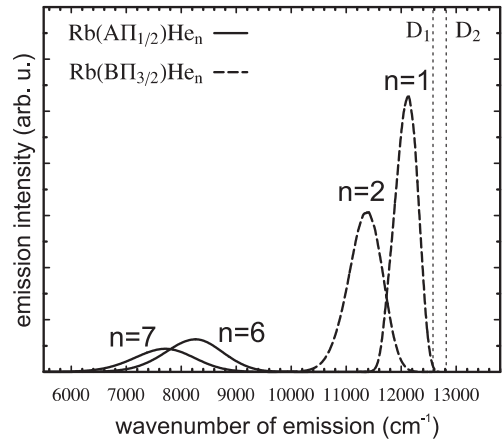}
\caption{Calculated emission spectra of Rb(B$^{2}\Pi_{3/2}$)He$_{n=1,2}$ (dashed lines) and Rb(A$^{2}\Pi_{1/2}$)He$_{n=6,7}$ (solid lines). The
dotted lines indicate the positions of the resonance lines of the free Rb atom.} \label{fig:EmissionSpectraTheo}
\end{figure}

As a last step we calculate the emission spectra $I(\nu)$ of all
Rb$^{\ast}$He$_{n=1\ldots9}$ exciplexes under the Franck-Condon
approximation as discussed in
\cite{CsHenExciplexesInSolidHe;Moroshkin}. The theoretical
emission spectra for Rb(B$^{2}\Pi_{3/2}$)He$_{n=1,2}$ and for
Rb(A$^{2}\Pi_{1/2}$)He$_{n=6,7}$ are shown in Fig.
\ref{fig:EmissionSpectraTheo}.

\section{Experimental results}\label{sect.ExperimentalResults}
\subsection{Experimental setup}\label{subsect:Setup}
The experimental setup is similar to the one described in our
previous publication
\cite{DiscoveryOfDumbbellShapedCsHenExciplexes;Nettels}. A helium
crystal is grown at pressures around 30 bar in a pressure cell
immersed in superfluid helium at 1.5 Kelvin. The matrix is doped
with rubidium atoms by laser ablation using a frequency-doubled
Nd:YAG laser. The cell has five windows for admitting the ablation
beam and the beam of the spectroscopy laser (a tunable cw
Ti:Al$_{2}$O$_{3}$ laser) and for collecting fluorescence from the
sample volume. The fluorescence is dispersed by a grating
spectrometer and recorded, depending on the spectral range under
investigation, either by a CCD camera
(9500$\dots$13500\,cm$^{-1}$) or by an InGaAs photodiode
(5500$\dots$9500\,cm$^{-1}$). We shall refer to these as
CCD-spectrometer and InGaAs-spectrometer respectively. With the
InGaAs-spectrometer spectra were recorded by a stepwise tuning of
the grating, while integral spectra could be recorded with the
CCD-spectrometer.

\subsection{Atomic Bubbles}\label{subsect:AtomicBubbles}
Defect atoms in solid helium reside in atomic bubbles, whose size
and structure can be described by the equilibrium between a
repulsive alkali-helium interaction due to the Pauli principle on
one hand and surface tension and pressure volume work on the other
hand
\cite{PressShiftResonanceLineBaLiHe;Kanorsky,{OpitcalPropAlkaliAtomsPressSuperfluidHe;Kinoshita},{PressureShiftOfAtomicResonanceLinesInLiquidSolidHe;Kanorsky}}.
The interaction with the helium bulk shifts the
$5S_{1/2}\rightarrow$5$P_{1/2}$ (D$_1$) and
$5S_{1/2}\rightarrow$5$P_{3/2}$ (D$_2$) transitions of Rb by
approximately 35\,nm to the blue with respect to their values
(794\,nm and 780\,nm respectively) in the free atom. This shift of
the excitation lines as well as a smaller blue shift of the
corresponding emission lines is well described by the bubble model
\cite{{PressureShiftOfAtomicResonanceLinesInLiquidSolidHe;Kanorsky},{LaserSpectroscopyOfAlkalineEarthAtomsInHeII;Bauer}}.
We have calculated the equilibrium radius of the atomic bubble
formed by the $5S_{1/2}$ ground state of the Rb atom to be
$R_b=6$\,\AA\ (Fig.\,\ref{fig:PotentialPlotRbHe2RbHe6}) following
the model described in
\cite{PressShiftResonanceLineBaLiHe;Kanorsky,OpitcalPropAlkaliAtomsPressSuperfluidHe;Kinoshita}.
For the interaction potential between groundstate Rb and He atoms
we have used the same semi-empirical potentials
\cite{LDependentPseudoPotentialAlkaliHe;Pascale} as for the
exciplex model.

It is the close vicinity of the helium atoms in the first
solvation shell, together with their large zero point oscillation
amplitudes, which form the basis of the efficient exciplex
formation in solid helium.

\subsection{Emission spectra following D$_{1}$ excitation}
\label{subsec.ObservedEmissionSpectraAfterD1Excitation}

\begin{figure}[h]
\includegraphics[width=8.5cm]{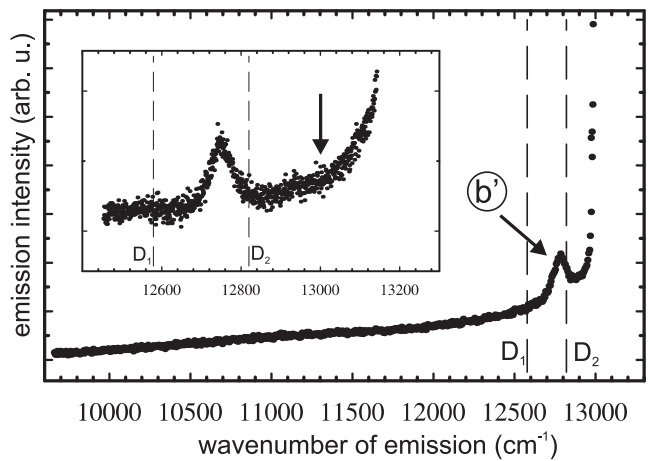}
\caption{Measured emission spectrum (dots) recorded with the
CCD-spectrometer following D$_{1}$ excitation. The dashed vertical
lines indicate the D$_1$ and D$_2$ lines of the free Rb atom. The
peak b' is the fluorescence from the D$_1$ transition. The inset
shows the spectral range around the D-lines recorded with a higher
resolution grating and an excitation frequency slightly
(65\,cm$^{-1}$) shifted to the blue. The rise on the right side is
from scattered laser light. The arrow gives the position at which
D$_{2}$ emission is detected after D$_{2}$ excitation (peak a in
Fig.\,\ref{fig:EmissionSpectrumD2Experiment}).}
\label{fig:EmissionSpectrumD1Experiment}
\end{figure}

Fig.\,\ref{fig:EmissionSpectrumD1Experiment} shows the emission
spectrum recorded with the CCD-spectrometer following excitation
at the D$_{1}$ wavelength 13140\,cm$^{-1}$ (758 nm). The peak at
12780\,cm$^{-1}$ represents fluorescence from the atomic
5$P_{1/2}$ state. While D$_{1}$ atomic fluorescence from Cs in
solid helium has been studied and used extensively in the past it
was believed that rubidium would not fluoresce on the D$_1$
transition when embedded in solid helium. This belief was based on
the reported quenching of that fluorescence at high pressures in
superfluid helium
\cite{PressureDependentQuenchingRb5PStatesLiquidHe;Yabuzaki}. It
should be noted that the observed Rb-D$_{1}$ fluorescence is
orders of magnitude weaker than the corresponding line in Cs and
could only be detected with long integration times (4 seconds) of
the CCD camera, which probably explains why this spectrum was not
observed in previous experiments
\cite{OptDetecNonradAlkaliAtoSolidHe;Eichler}.

The apple-shaped exciplexes with one or two bound helium atoms are
expected to fluoresce within the spectral range of
Fig.\,\ref{fig:EmissionSpectrumD1Experiment} and the absence of
any prominent spectral feature indicates that these complexes are
not formed upon D$_{1}$ excitation. The sloped background visible
in Figs.\,\ref{fig:EmissionSpectrumD1Experiment} and
\ref{fig:EmissionSpectrumD2Experiment} is a strong wing of
scattered laser light. The inset in
Fig.\,\ref{fig:EmissionSpectrumD1Experiment} shows a spectrum
which was recorded using a grating with a higher resolution. The
excitation laser was shifted by 65\,cm$^{-1}$ (still in the
D$_{1}$ absorption band (Fig.\,\ref{fig.ExcitationSpectrumRbD1D2})
to the blue with respect to the spectrum of
Fig.\,\ref{fig:EmissionSpectrumD1Experiment} to make clear that no
D$_{2}$ emission can be observed after D$_{1}$ excitation. The
arrow in the inset indicates the position of the D$_{2}$ emission
measured after D$_{2}$ exciation (peak a in
Fig.\,\ref{fig:EmissionSpectrumD2Experiment}).

\begin{figure}[h]
\includegraphics[width=8.5cm]{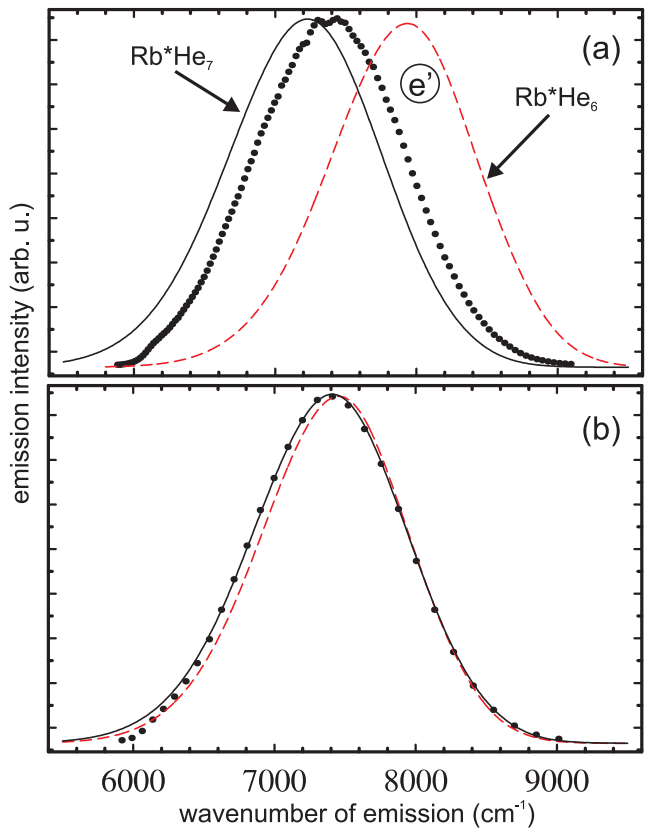}
\caption{Fluorescence spectrum (dots) following D$_1$ excitation
measured with the InGaAs-spectrometer. The emission band stems
from a Rb$^{\ast}$He$_{n>2}$ exciplex (e'). An identical emission
spectrum was observed after D$_2$ excitation. (a) The dashed line
is a calculated emission spectrum from Rb$^{\ast}$He$_{6}$ and the
solid line from Rb$^{\ast}$He$_{7}$. (b) The two theoretical
spectra are shifted in order to match the experimental curve.}
\label{fig:RbHe6EmissionD1Excitation}
\end{figure}

When exploring the longer wavelength range with the
InGaAs-spectrometer we found a very strong fluorescence band
(Fig.\,\ref{fig:RbHe6EmissionD1Excitation}) centered at
7420\,cm$^{-1}$, which we assign to Rb$^{\ast}$He$_{n>2}$
exciplexes in the A$^{2}\Pi_{1/2}$ state. This is the first
recording of such exciplexes after D$_{1}$ excitation and the
proof that the quenching of atomic D$_{1}$ fluorescence is due to
exciplex formation. The dashed and the solid lines in
Fig.\,\ref{fig:RbHe6EmissionD1Excitation} are theoretical emission
spectra from Rb$^{\ast}$He$_{6}$ and Rb$^{\ast}$He$_{7}$
respectively. Figure\,\ref{fig:RbHe6EmissionD1Excitation}(b) shows
the theoretical curves, shifted such as to make their blue wings
coincide with the experimental points. The line shape of the
experimental curve is well reproduced by the two theoretical
curves. The theoretical curve of the Rb$^{\ast}$He$_{7}$ fits the
experimental points better on the low energy side, while on the
high energy range both curves are in very good agreement with the
experimental spectrum. A small discrepancy is visible on the low
energy wing, which can be due to imprecisions of the strongly
sloped ground state potential
(Fig.\,\ref{fig:PotentialPlotRbHe2RbHe6}) or to changes of the
latter due to the helium bulk. It is a remarkable fact that the
fluorescence yield of this exciplex after D$_{1}$ excitation in
solid helium is larger than after D$_{2}$ excitation, while it was
not observed at all in superfluid helium. We will come back to
this point in Sect.\,\ref{sect.ExperimentalResults}.

A similar emission at around 7200\, cm$^{-1}$ has been seen in
liquid helium by the Kyoto group
\cite{EmissionSpectraRbHeExciplexesColdHeGas;Yabuzaki} after
D$_{2}$ excitation and was assigned to the emission by the
Rb$^{\ast}$He$_{6}$ exciplex.


\subsection{Emission spectra following D$_{2}$ excitation}\label{subsec.ObservedEmissionSpectraAfterD2Excitation}

\begin{figure}[h]
\includegraphics[width=8.5cm]{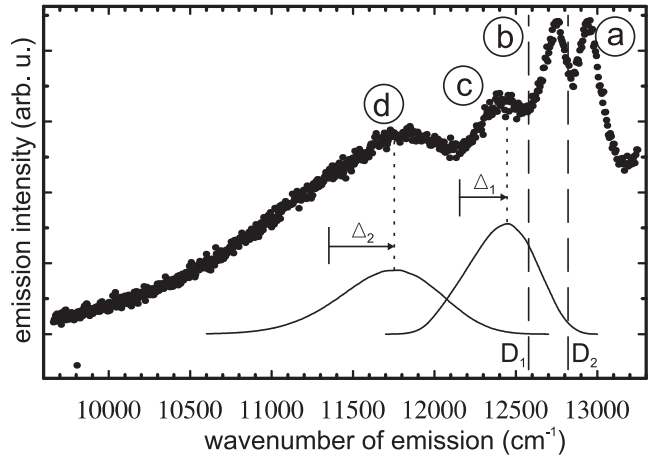}
\caption{Fluorescence spectrum (dots) recorded with the
CCD-spectrometer following D$_{2}$ excitation. The dashed vertical
lines indicate the position of the D$_{1}$ and D$_{2}$ line of the
free Rb atom. The following assignments are made to the emission
peaks: atomic D$_{2}$ fluorescence (a), atomic D$_{1}$
fluorescence (b), emission from Rb(B$^{2}\Pi_{3/2}$)He$_{1}$
exciplexes (c), and emission from Rb(B$^{2}\Pi_{3/2}$)He$_{2}$
exciplexes (d). The solid lines are calculated emission spectra
from Rb(A$^{2}\Pi_{3/2}$)He$_{1}$ and Rb(A$^{2}\Pi_{3/2}$)He$_{2}$
exciplexes. The lines are shifted in order to match the peaks of
the experimental curves. $\Delta_{1}=350$\,cm$^{-1}$ and
$\Delta_{2}=440$\,cm$^{-1}$ are the shifts with respect to the
calculated positions shown in
Fig.\,\ref{fig:EmissionSpectraTheo}.}
\label{fig:EmissionSpectrumD2Experiment}
\end{figure}

Fig.\,\ref{fig:EmissionSpectrumD2Experiment} shows the emission
spectrum, measured with the CCD-spectrometer, when the laser is
tuned to the atomic D$_{2}$ transition at 13420\,cm$^{-1}$ (745
nm).

Four prominent spectral features can be seen in the emission
spectrum. The two rightmost peaks (labelled a and b) represent
atomic D$_{2}$ and D$_{1}$ fluorescence respectively. Together
with the peak of Fig.\,\ref{fig:EmissionSpectrumD1Experiment} they
constitute the first observation of atomic fluorescence from
rubidium in solid helium. The presence of D$_{1}$ emission after
D$_{2}$ excitation is evidence for the existence of a fine
structure relaxation channel. We assign the two broader features c
and d peaked at 12400\,cm$^{-1}$ and 11800\,cm$^{-1}$ respectively
to the emission from Rb(B$^{2}\Pi_{3/2}$)He$_{1}$ and
Rb(B$^{2}\Pi_{3/2}$)He$_{2}$ exciplexes. The solid lines in
Fig.\,\ref{fig:EmissionSpectrumD2Experiment} are the calculated
$n=1$ and $n=2$ emission spectra of
Fig.\,\ref{fig:EmissionSpectraTheo} shifted to the blue by
$\Delta_1$ and $\Delta_2$ respectively, so that their line centers
coincide with the positions of the measured curves. The shifts are
probably due to the interaction with the surrounding helium
bubble. Note that the two theoretical curves have to be shifted by
different amounts in order to match the experimental lines. We
have found previously in the Cs-He system
\cite{CsHenExciplexesInSolidHe;Moroshkin} that the rate and sign
of the pressure shift of exciplex emission lines depend on the
number of bound helium atoms.

As with the spectra of
Sect.\,\ref{subsec.ObservedEmissionSpectraAfterD1Excitation} we
have recorded the emission in the region of longer wavelengths
with the InGaAs-spectrometer. As a result we find a spectrum,
which is identical (same central wavelength and same width) with
the one observed with D$_{1}$ excitation
(Fig.\,\ref{fig:RbHe6EmissionD1Excitation}). This suggests that
the emission stems from the same state (A$^{2}\Pi_{1/2}$) as the
emission after D$_{1}$ excitation. The population of that state
following D$_{2}$ excitation is another proof of the existence of
a fine structure relaxation mechanism. No other exciplex emission
was observed in the spectral range between the
Rb$^{\ast}$He$_{n>2}$ and the Rb$^{\ast}$He$_{2}$ exciplexe
emission (peak e' in Fig.\,\ref{fig:RbHe6EmissionD1Excitation}(a)
and peak d in Fig.\,\ref{fig:EmissionSpectrumD2Experiment}
respectively).

\subsection{Atomic excitation spectra}\label{subsec.AtomicExcitationSpectra}
The experimental emission spectra presented above were recorded
with two fixed excitation wavelengths, chosen such as to maximize
the signals of interest. It is of course interesting to
investigate how the different spectral features depend on the
excitation wavelength. For this we have varied the wavelength of
the Ti:Al$_{2}$O$_{3}$ laser in discrete steps  over the spectral
range of 13000$\dots$13700\,cm$^{-1}$ ($\sim$770$\dots$730\,nm).
For every excitation wavelength we have measured the amplitudes of
the emission peaks of Figs.
\ref{fig:EmissionSpectrumD1Experiment},
\ref{fig:RbHe6EmissionD1Excitation} and
\ref{fig:EmissionSpectrumD2Experiment}.

\begin{figure}[h]
\includegraphics[width=6.5cm]{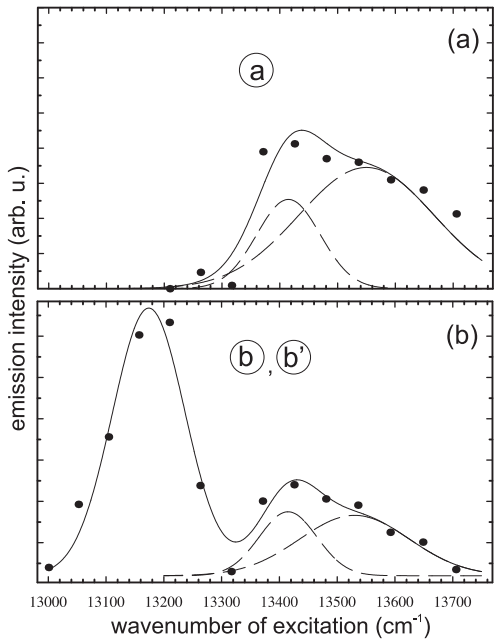}
\caption{Excitation spectra of the fluorescence from atomic
rubidium: Top: fluorescence analyzing spectrometer set to the
D$_{2}$ emission line (peak a of
Fig.\,\ref{fig:EmissionSpectrumD2Experiment}); bottom:
spectrometer set to the D$_1$ emission line (peaks b' and b of
Figs.\,\ref{fig:EmissionSpectrumD1Experiment} and
\ref{fig:EmissionSpectrumD2Experiment}). The dashed lines are
Gaussians whose sum (solid line) was fitted to the data.}
\label{fig.ExcitationSpectrumRbD1D2}
\end{figure}

The top part of Fig.\,\ref{fig.ExcitationSpectrumRbD1D2} shows the
excitation spectrum of D$_{2}$ fluorescence, which is centered at
13460\,cm$^{-1}$ (743 nm). One sees clearly that this fluorescence
can only be produced by D$_2$ excitation. The lower part of
Fig.\,\ref{fig.ExcitationSpectrumRbD1D2} shows the excitation
spectrum of D$_1$ fluorescence. It consists of two absorption
bands centered at 13180\,cm$^{-1}$ and 13460\,cm$^{-1}$
respectively, which corresponds to excited states correlating with
the atomic 5$P_{1/2}$ and 5$P_{3/2}$ levels respectively. D$_1$
fluorescence can thus be produced directly via D$_1$ excitation or
via D$_2$ excitation combined with a $J$-mixing interaction due to
the alkali-helium interaction.
The D$_1$ absorption band is slightly asymmetric with a longer
wing on the low energy side. This feature has been observed before
in Cs
\cite{PressureShiftOfAtomicResonanceLinesInLiquidSolidHe;Kanorsky}.
The  D$_2$ absorption band measured for both D$_1$ and D$_2$
fluorescence, has a double peaked-structure. The scarce number of
data points is well fitted by a superposition of two Gaussians
separated by about 125\,cm$^{-1}$. This splitting of the D$_2$
excitation lines of cesium and rubidium in superfluid helium has
been explained before in terms of a dynamic Jahn-Teller effect due
to quadrupolar bubble-shape oscillations which lift the degeneracy
of the $P_{3/2}$ state
\cite{DoublyShapedD2ExcSpecCsRbinSupfluHeQuadruBubbleOsci;Yabuzaki}.

\subsection{Exciplex excitation spectra}\label{subsec.AtomicExcitationSpectra}
\begin{figure}[h]
\includegraphics[width=6.5cm]{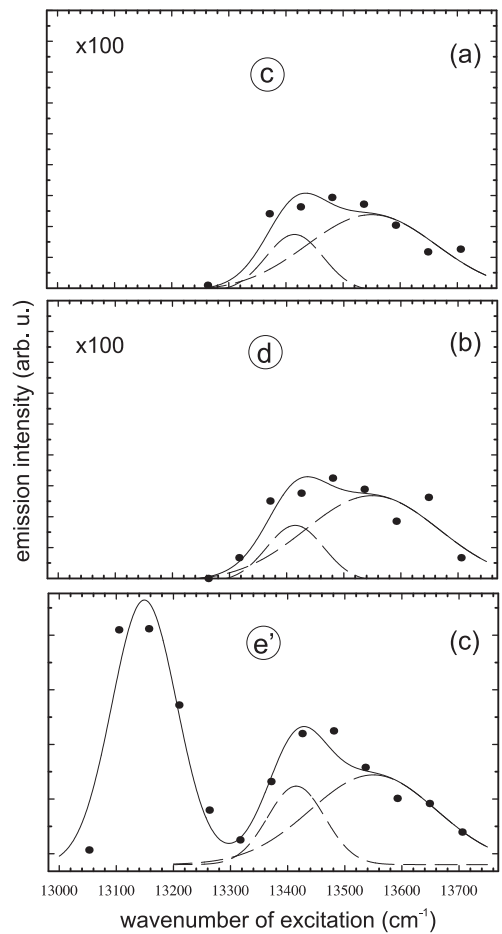}
\caption{Excitation spectra of the fluorescence from
Rb$^{\ast}$He$_{n}$ exciplexes (dots) with the fluorescence
spectrometer tuned to emission from Rb$^{\ast}$He$_{1}$ c,
Rb$^{\ast}$He$_{2}$ d, and Rb$^{\ast}$He$_{n_{max}}$ e'. The solid
lines are Gaussian fits. The signals in the spectrum e' is
approximately two orders of magnitude larger than the ones of c
and d and than the atomic signals from
Fig.\,\ref{fig.ExcitationSpectrumRbD1D2}.}
\label{fig.ExcitationSpectrumRbHe1RbHe2RbHe6}
\end{figure}
Fig.\,\ref{fig.ExcitationSpectrumRbHe1RbHe2RbHe6} shows the
excitation spectra of the exciplex lines c, d, and e' of
Figs.\,\ref{fig:RbHe6EmissionD1Excitation} and
\ref{fig:EmissionSpectrumD2Experiment}. As the
Rb$^{\ast}$He$_{1,2}$ exciplexes can only be observed after D$_2$
excitation (Fig.\,\ref{fig.ExcitationSpectrumRbHe1RbHe2RbHe6} c,
d) we conclude that these apple-shaped complexes are formed in the
B$^{2}\Pi_{3/2}$ state. The D$_1$, D$_2$ and
Rb(B$^{2}\Pi_{3/2}$)He$_{1,2}$ emission lines are very weak and of
similar amplitude. The bottom spectrum (e') represents by far the
strongest signal that comes from the
Rb(A$^{2}\Pi_{1/2}$)He$_{n>2}$ exciplex which can be excited by
either D$_1$ or D$_2$ radiation. Its emission line is about 100
times stronger than the other lines. This result is in strong
contrast with the emission of the corresponding cesium exciplex,
Cs(A$^{2}\Pi_{1/2}$)He$_{n>2}$, in solid helium, for which the
emission after D$_1$ excitation is very weak
\cite{CsHenExciplexesInSolidHe;Moroshkin}. The double-peaked
structure of the D$_2$ excitation spectrum is not well resolved
for the Rb$^{\ast}$He$_{1,2}$ exciplexes. It was observed before
for Cs$^{\ast}$He and Rb$^{\ast}$He exciplexes on superfluid
helium nanodroplets
\cite{SpectroscopyOfCsAttachedToHeliumNanodroplets;Stienkemeier,RbHeExciplexFormationOnHeliumNanodroplets;Ernst}.

\section{Discussion}\label{sect.Discussion}
\subsection{Atomic lines}\label{subsect.AtomicLines}

The assignment of the atomic D$_1$ and D$_2$ excitation and
emission lines is unambiguous. The excitation lines are
blue-shifted by approximately 600\,cm$^{-1}$, while the emission
lines are shifted by only 65\,cm$^{-1}$ with respect to the free
atomic transitions. These shifts have been studied in superfluid
helium \cite{OpitcalPropAlkaliAtomsPressSuperfluidHe;Kinoshita}
and are well described by the bubble model. The blue shift results
from the interaction with the bulk helium, which is less
pronounced in the emission process as the latter occurs in a
bubble of larger size
\cite{PressureShiftOfAtomicResonanceLinesInLiquidSolidHe;Kanorsky}.
As already mentioned, excitation at the D$_1$ transition leads to
emission on the D$_1$ line only, while excitation at the D$_2$
line leads to emission on both the D$_1$ and the D$_2$ lines.

\subsection{Apple-shaped Rb(B$^{2}\Pi_{3/2}$)He$_{1,2}$ exciplexes}\label{subsect.RbHe12}

One or two helium atoms approaching the apple-shaped atomic
5P$_{3/2}$, m$_J=\pm$3/2 state do not experience a potential
barrier on their way to the potential well of the B$^{2}\Pi_{3/2}$
state. The formation process of Rb$^{\ast}$He$_{1}$ and
Rb$^{\ast}$He$_{2}$ exciplexes is therefore straightforward after
D$_2$ excitation. Note that the potential diagram for
Rb$^{\ast}$He$_{1}$ is similar to the one for Rb$^{\ast}$He$_{2}$,
shown in Fig.\,\ref{fig:PotentialPlotRbHe2RbHe6}, and that it has
a reduced potential well depth. The Rb$^{\ast}$He$_{1,2}$ exciplex
emission line following D$_1$ excitation is not observed because
only the largest exciplex is formed as discussed in paragraph
\ref{subsect.FormationProcessRbHe7}.

Cs$^{\ast}$He$_{2}$ is the only apple-shaped exciplex that was
observed in related experiments with cesium in superfluid
\cite{EmissionSpectraCsHeExcimersColdHeGas;Yabuzaki} and in solid
\cite{CsHenExciplexesInSolidHe;Moroshkin,DiscoveryOfDumbbellShapedCsHenExciplexes;Nettels}
helium, while in cold helium gas both Cs$^{\ast}$He$_{1}$ and
Cs$^{\ast}$He$_{2}$ structures were detected
\cite{EmissionSpectraCsHeExcimersColdHeGas;Yabuzaki}. It remains
an open question why the Cs$^{\ast}$He$_{1}$ exiplex does not
fluoresce in condensed helium, while the corresponding rubidium
exciplex does.

\subsection{Dumbbell-shaped Rb(A$^{2}\Pi_{1/2}$)He$_{n>2}$ exciplexes}\label{subsect.RbHe67}

The emission line shown in
Fig.\,\ref{fig:RbHe6EmissionD1Excitation} has the longest
wavelength of all observed spectral lines and originates thus from
the lowest-lying bound state, i.e., the A$^{2}\Pi_{1/2}$ state of
Fig.\,\ref{fig:PotentialPlotRbHe2RbHe6}(b). Note that all
Rb$^{\ast}$He$_{n>2}$ exciplexes have similar potential curves
with potential wells/barriers increasing with $n$. All of these
structures have the shape of dumbbells, with the helium atoms
bound around their waists
\cite{CsHenExciplexesInSolidHe;Moroshkin}.
Fig.\,\ref{fig:RbHe6EmissionD1Excitation} also shows the
calculated line shapes of the emission from Rb$^{\ast}$He$_{6}$
and Rb$^{\ast}$He$_{7}$. Disregarding shifts of the line centers
the theoretical line shapes match the experimental spectrum quite
well. The good matching of the line width in particular indicates
that this emission is from a single exciplex species with a
specific number of bound helium atoms and that it does not come
from a superposition of different exciplexes. The shift of the
lines is most likely due to the interaction with the helium bulk,
which was not taken into account in our calculation. It is
difficult to estimate whether the bulk shifts the line to the blue
or to the red. One can therefore not assign the observed emission
band to Rb$^{\ast}$He$_{6}$ or Rb$^{\ast}$He$_{7}$ in an
unambiguous way. The calculated binding energies
$\epsilon_{b}$(Rb) (Fig.\,\ref{fig:EnergyLevel}) show that the
complex with 6 helium atoms has the lowest binding energy and is
therefore the most stable exciplex. Observations in liquid He
\cite{EmissionSpectraRbHeExciplexesColdHeGas;Yabuzaki} confirm
this prediction. However, the exact calculation of the energy of
the lowest lying bound state involves a precise quantitative
treatment of its oscillatory degrees of freedom. In
\cite{CsHenExciplexesInSolidHe;Moroshkin} we have described in
detail how we calculate these oscillation energies. There is an
uncertainty in the calculated binding energies due to the
simplified assumptions we made. An additional uncertainty comes
from the semi-empirical pair potentials
\cite{LDependentPseudoPotentialAlkaliHe;Pascale}. For big
exciplexes like the Rb$^{\ast}$He$_{6}$ every uncertainty in the
potential will be amplified because of the additive contribution
of the $n$ helium atoms discussed in Sect.\,\ref{sect.Theory}.
This can change the position and the depth of the well in the
excited state. To all of this adds the effect of the helium bulk,
which was not treated so far. The following arguments support the
Rb$^{\ast}$He$_{6}$ to be the structure observed. It has the
minimal binding energy and the corresponding Cs exciplex line is
shifted to lower wavenumbers with increasing pressure
\cite{CsHenExciplexesInSolidHe;Moroshkin}. Assuming the same
tendency for the Rb exciplex brings the spectral position of
Rb$^{\ast}$He$_{6}$ into better agreement with the experimental
curve (Fig.\,\ref{fig:RbHe6EmissionD1Excitation}). On the other
hand the line shape of the calculated Rb$^{\ast}$He$_{7}$ fits
better to the data. Therefore we can not conclude which exciplex
is the one observed in the experiment.

\subsection{Formation of dumbbell-shaped Rb(A$^{2}\Pi_{1/2}$)He$_{n>2}$ exciplexes}\label{subsect.FormationProcessRbHe7}

The radius of the bubble formed by the rubidium ground state has
an equilibrium radius R$_{b}$ of $6$\,\AA, which is smaller than
the corresponding radius for cesium. The excitation process is a
Franck-Condon transition to the $5P$ state during which the radius
does not change.

The D$_1$ excitation starting at $R_{b}$(5S)$=6$\,\AA\, ends at
the left of the potential barrier of the A$^{2}\Pi_{1/2}$ state so
that the exciplex is easily formed by helium atoms dropping into
the well. Note that for cesium in solid helium the corresponding
transition ends on the right side of the potential barrier in the
excited state \cite{CsHenExciplexesInSolidHe;Moroshkin}. In that
case the helium atoms have to tunnel through the potential barrier
in order to form the exciplex. This explains why  exciplex
emission of Cs in solid helium after D$_1$ excitation is much
weaker than after D$_2$ excitation, while for Rb the opposite
holds. It also explains why no emission from Rb exciplexes after
D$_1$ excitation could be observed in gaseous (below 10 Kelvin)
and in liquid helium environments
\cite{EmissionSpectraRbHeExciplexesColdHeGas;Yabuzaki} in which
the helium atoms are, on average, further away from the Rb atom
and where the excitation thus ends at the right of the potential
barrier. Under those conditions the exciplex formation is strongly
suppressed as the helium atoms have to tunnel one after another
through the potential barrier to form the exciplex. This tunneling
occurs at a rate which is smaller than the exciplex lifetime. The
same is true for Rb on He droplets, where no exciplex was observed
after D$_{1}$ excitation
\cite{AlkaliHeliumExciplexFormationOnTheSurfaceOfHeliumNanodroplets2;Scoles}.
The authors of
\cite{AlkaliHeliumExciplexFormationOnTheSurfaceOfHeliumNanodroplets2;Scoles}
estimated the tunneling time to be about 500\,ns, much longer than
the lifetime.

When exciting the system at $R_{b}$(5S)$=6$\,\AA\, on the D$_2$
transition the corresponding fine-structure relaxation channel
allows the system to form the terminal exciplex in the potential
well of the A$^{2}\Pi_{1/2}$ state.

In solid helium only the largest exciplex
Rb$^{\ast}$He$_{n_{max}}$ is observed after D$_1$ excitation. This
means that the potential well is filled up to the maximal value of
helium atoms that it can hold on a time scale which is shorter
than the radiative lifetimes of the intermediate products. It is
therefore likely to assume, as we have previously done for the
formation of the corresponding cesium exciplexes that the exciplex
results from a collective motion of the helium atoms.


\subsection{Summary and conclusion}
We have presented several new spectral features observed in the
laser-induced fluorescence from a helium crystal doped by laser
ablation from a solid rubidium target. We detected for the first
time weak, but unambiguously identified D$_1$ and D$_2$
fluorescence lines from atomic rubidium, which were previously
believed to be completely quenched in solid helium. We have shown
that Rb$^*$He$_n$ exciplex formation is possible after D$_1$
excitation, in contrast to cesium doped He, in which exciplex
formation proceeds only via absorption on the D$_2$ transition. We
have explained this in terms of the smaller bubble diameter of
rubidium, which allows the excitation to proceed directly to a
binding state without tunnelling processes as they are needed with
cesium. We have further reported the observation of
Rb$^{\ast}$He$_{1,2}$ exciplex emission after D$_2$ excitation, a
process which could not be observed in liquid helium, as well as
the observation of a larger exciplex. The main decay channel of
laser excited Rb in solid helium is via the formation of this
largest exciplex, assigned to be either Rb$^{\ast}$He$_{6}$ or
Rb$^{\ast}$He$_{7}$ with subsequent emission of strongly red
shifted fluorescence.

It remains an open question why one observes the two exciplexes
Rb$^{\ast}$He$_{1}$ and Rb$^{\ast}$He$_{2}$, while in equivalent
experiment with cesium there is only fluorescence form the
Cs$^{\ast}$He$_{2}$ complex. This feature could be related to a
recently discovered absorption band of the Rb$_{2}$ dimer which
overlaps with the D$_2$ atomic absorption line
\cite{SpectroscopyOfAlkaliDimersInSolidHe;Moroshkin}. It may also
be related to unexplained details of the different steps in the
formation process. Femtosecond pump-probe experiments would
clearly be the tool of choice for further investigations of this
question.

\begin{acknowledgments}
We thank J. Pascale for sending us his numerical Rb-He pair
potentials. This work was supported by the grant number
200020-103864 of the Schweizerischer Nationalfonds.
\end{acknowledgments}



\bibliography{PaperListe}

\end{document}